# Transport critical current of Solenoidal MgB$_2$/Cu Coils Fabricated Using a Wind-Reaction In-situ Technique

S. Soltanian, J. Horvat, X.L. Wang, M. Tomsic* and S.X. Dou

Institute for Superconducting and Electronic Materials
University of Wollongong, Northfields Ave., Wollongong NSW 2522 Australia
*Hyper Tech Research Inc, 110E. Canal St. Troy, OH 45373

**Abstract**

In this letter, we report the results of transport $J_c$ of solenoid coils upto 100 turns fabricated with Cu-sheathed MgB$_2$ wires using a wind-reaction in-situ technique. Despite the low density of single core and some reaction between Mg and Cu-sheath, our results demonstrate the decrease in transport $J_c$ with increasing length of MgB$_2$ wires is insignificant. Solenoid coils with diameter as small as 10 mm can be readily fabricated using a wind-reaction in-situ technique. The $J_c$ of coils is essentially the same as in the form of straight wires. A $J_c$ of 133,000 A/cm$^2$ and 125,000 A/cm$^2$ at 4K and self field has been achieved for a small coil wound using Cu-sheathed tape and Cu-sheathed wire respectively. These results indicate that the MgB$_2$ wires have a great potential for lage scale applications

Since the discovery of the 39K superconductor, MgB$_2$ [1], a significant advancement has been achieved in fabrication of various forms of MgB$_2$. In particular, intensive efforts have been made in improving the critical current density ($J_c$) in various metal sheathed MgB$_2$ wires. High $J_c$ of $10^5 - 10^6$ A/cm$^2$ at 4 K to 30 K for MgB$_2$ wires and good performance of $J_c$ in magnetic field have been reported by several groups [2-9]. However, the results reported thus far have been largely limited to the short samples of several centimetres long. In contrast, long Bi-based HTS wires with high $J_c$ were reported in several months time after discovery of the Bi-HTS compound [10]. For large scale applications it is essential to fabricate this material into long wires and coils. The critical challenge remained is how much the $J_c$ deteriorates with increasing length of the wire and whether one can wind the wire into a coil without appreciable loss of $J_c$. In this letter, we report the fabrication and transport critical current of solenoidal MgB$_2$/Cu coils fabricated using a wind-reaction in-situ technique.

Standard powder-in-tube methods were used for the Cu-Fe or Cu clad MgB$_2$ tape. The pure Cu or Fe tubes had an outside diameter (OD) of 10 mm, a wall thickness of 1 mm, and was 10 cm long. One end of the tube was sealed, and the tube was filled in with magnesium (99% purity) and amorphous boron (99%) with the stoichiometry of MgB$_2$. The remaining end was crimped by hand. The composite was drawn to a 0.5, 0.7 and 1 mm diameter rod several meters long. Some wires were further rolled to ribbon over many steps. Several short samples 2 cm in length were cut from the wire and ribbon. The green wires were wound to ceramic tube of 8mm diameter with both ends fixed to the slots at the end of ceramic tube. Several coils were prepared using this procedure, and one of the coils has 100 turns wound using 3 meter long Cu-sheathed MgB$_2$ wire. These coils and some straight pieces of wires and tapes were then sintered in a tube furnace over a temperature range from 650-800$^o$C for 10 min. A high purity argon gas flow was maintained throughout the sintering process.

Short pieces of wires at 4mm and 30mm were used for magnetic and transport critical current measurements using four probe. For longer wire samples up to 300 mm, a number of voltage contacts were made at different distance to determine the variation of $I_c$ with length. For the 100 turn coil two voltage contacts were made at a distance 2 meter apart with current contacts at the end of the coil. Transport measurements of the voltage versus current (V-I) were performed by a standard 4-probe, DC method. The measurements were performed in liquid helium. Current contacts were soldered on the samples at least 1cm away from the voltage contacts, to allow for the heat created at the current contacts to dissipate into liquid helium before it reaches the part of the sample between the voltage contacts. The current was switched on and off gradually, to avoid damage of the sample due to the mechanical shock upon fast change of the current. Each point in V-I was taken within 10 seconds, to avoid heating of the sample. Magnetic measurements were performed by a Quantum Design PPMS magnetometer. Magnetic hysteresis loops were used to obtain the field dependence of the critical current density at each of the temperatures measured, by employing the critical state model. The sweep rate of the field was 50 Oe/s. Magnetic field was always applied along the sample.

Figure 1 displays the 10mm diameter $MgB_2$ coils of 100 turns, wound using 3 meter Cu-sheathed single core wire and heat-treated at $750^o$C for 10min. The coils are reasonably flexible and can be stretched within 20% distance and bent to an angle of $30^o$ after heat treatment without degradation in $I_c$. Figure 2 shows the photomicrographs of the transverse and longitudinal cross section of the 100 turn solenoid coil (Coil-4). It is seen that the interface between Cu and $MgB_2$ core is very smooth. Although there is thin layer of MgCu formed during sintering the reaction between Cu and Mg is not very serious as a short reaction time (10min) and low sintering temperatures ($750^0$C) were used.

Table 1 lists the parameters for samples in the forms of straight wires, coils with Cu-sheath. These results show very interesting features. Compare wire-1 and coil-1 which is made from the same green wire, the $J_c$ is the same for both cases, indicating that (a) winding the wire to 10mm diameter coil does not degrade $J_c$ and (b) the increase of distance between two voltage contacts from 13mm to 100mm did not cause a reduction in $J_c$. The $J_c$ results obtained from a straight wire (wire-2) and 100 turn coil (Coil-4) further indicate that there is no evidence of significant length dependence of $J_c$. For the sample wire-2, the critical current was not reached due to the limit of contact heating. But the $I_c$ for all three contact distances exceeds 100A. The 6 turn coil (Coil-2) and 5 turn coil (Coil-3) were wound using thinner Cu-sheathed wire (OD = 0.5mm) and tape respectively. Because their core density is relatively higher a $J_c$ of 125,000 A/cm$^2$ and 133,000 A/cm$^2$ at helium temperature and self field has been achieved respectively, suggesting that density is one of the most critical factors to influence $J_c$.

Figure 3 shows the characteristics of the voltage – current curves (V-I) for the 100 turn solenoid coil with two voltage contacts at distance of 35 mm (1 turn) and 2 meters (63 turns) apart. Despite the possible inhomogeneity along the length and low density of the core, the same value of $J_c$ for the two large different voltage contact distances clearly indicates that the deterioration of $J_c$ with increasing length of $MgB_2$ wires is insignificant. It should be pointed out that the measurement for the voltage contact distance of 35mm was performed after several cycle of the coil between helium temperature and room temperature due to the requirement of change of contact position. It is expected that the cycling could cause some degradation in $J_c$ as the density of the single core is low. Furthermore, the solenoid coil with 100 turns passing 73 A will generate a self-field of about 80mT which will slightly reduce the $I_c$. So, the $J_c$ for 35mm contact distance should be higher than that of 2 meters contact distance. Compared to sample Wire-1 which was made from the same batch of the green wire the $J_c$ for the 35mm contact distance would not be higher than that of Wire-1. If the density of the wire core can be improved the thermal cycling would not have detrimental effect as has been demonstrated in Fe-sheathed wire [12].

It is also noted from figure 3 that the V-I curve for the contact distance of 35mm shows a steep increase at the critical current 72A while the voltage for the large contact distance increases more gradually. For the former the V-I characteristics beyond the critical current is dominated by the Cu sheath while in the latter case, there are still large segments of the coil remaining superconducting when a small section becomes normal. It is unclear that what kind of voltage criteria should be used to determine the $I_c$ for the Cu-sheathed $MgB_2$ wire at this stage. In our measurements we use the same standard to determine the $I_c$ for short straight wire and for the solenoid coils. This means that criterion for the 100 turn solenoid coil is about 0.005μV/cm, compared with common criterion of 1μV/cm used in HTS wires. If we apply the 1μV/cm criterion to the 100 turn coil the $I_c$ will be about 200A, estimated by extrapolating the V-I curve. This will cause huge heating over normal sections of coil. Thus, the 1μV/cm criterion may be applicable to the metal sheathed $MgB_2$ wire.

In summary, it is evident that long $MgB_2$ wires and solenoid coils can be fabricated using wind-reaction in-situ technique with little $J_c$ degradation over the entire length, paving the way for design and fabrication of magnetic windings and magnets which are central element for many large scale applications.

Acknowledgments:
We are grateful to the helpful discussion with Drs T. Silver, M. Ionescu, A. Pan, M.J. Qin, H.K. Liu, M. Sumption and E.W Collings. This work is supported by the Australian Research Concil and Industry partner, Hyper Tech Research Inc at Troy OH USA.

Figure captions:

Figure 1. The appearance of two 10mm diameter $MgB_2$ coils of 100 turns, wound using 3 meter Cu-sheathed single core wire and a 10mm diameter coil of 10 turns. These coils were heat-treated at 750°C for 10min.

Figure 2. The photomicrographs of the transverse (a) and longitudinal cross section (b) of the 100 turn coil-4. The scale bar in Fig 2 represents 300μm.

Figure 3. The voltage – current curves (V-I) for the 100 turn solenoid coil with two voltage contacts at distance of 35 mm (1 turn) and 2 meters (63 turns). The two current contacts were soldered at the end of each side of the solenoid coil for the latter case. After this measurement two voltage contacts were soldered at the center of coil at a distance of 35mm apart while current contacts were moved to one turn away from the voltage contacts.

Table 1. List of various samples with description and measurement results of $J_c$

| Designation of samples | Sample description | Distance between voltage contacts (D, mm) | $MgB_2$ Core diameter (mm) | $I_c$ (A) | $J_c$ (A/cm$^2$) |
|---|---|---|---|---|---|
| Wire-1 | 30mm $MgB_2$/Cu straight wire | 13 | 0.45 | 105 | 66,000 |
| Wire-2 | 200mm long $MgB_2$/Fe-Cu straight wire | 37 | 0.6 | >100 | >35,300 |
| | | 80 | 0.6 | >100 | >35,300 |
| | | 120 | 0.6 | >100 | >35,300 |
| Wire Coil-1 | 5 turn $MgB_2$/Cu coil, 10mm OD | 100 (3 turns) | 0.45 | >101 | >63,500 |
| Wire Coil-2 | 6 turn $MgB_2$/Cu coil, 10mm OD | 140 (4 turns) | 0.32 | 100 | 125,000 |
| Tape Coil-3 | 5 turn $MgB_2$/Cu coil, 10mm OD | 100 (3 turns) | 0.055 | >73 | >133,000 |
| Wire Coil-4 | 100 Turn $MgB_2$/Cu solenoid, 10mm OD | 35 (1 turn) | 0.45 | 72 | 45,300 |
| | | 2000 (63 turns) | 0.45 | 73 | 45,900 |

Fig 1

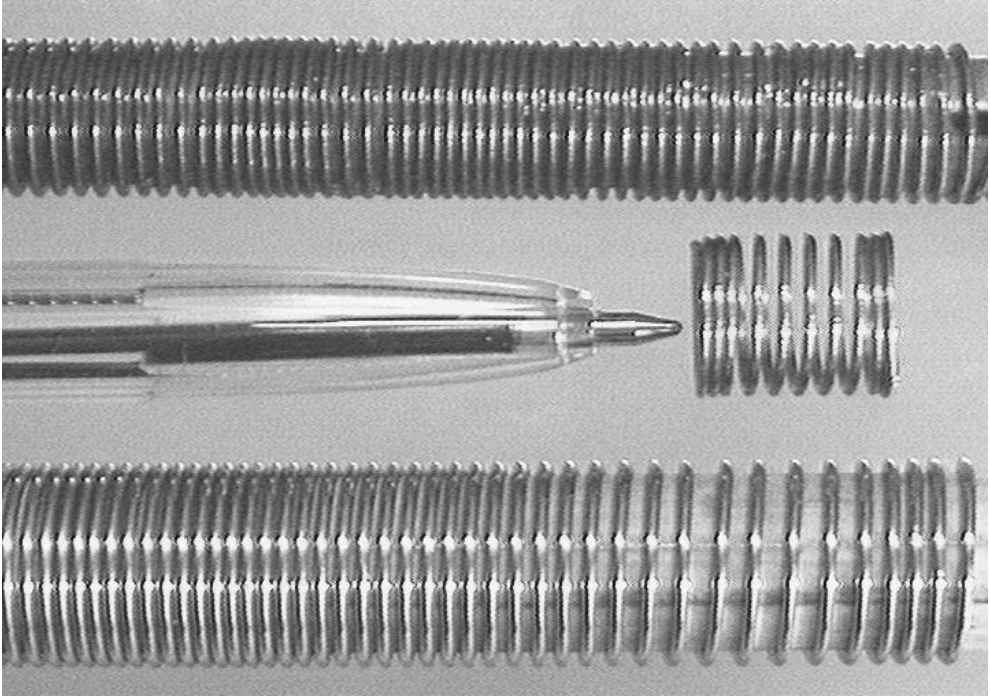

Fig 2 (a)

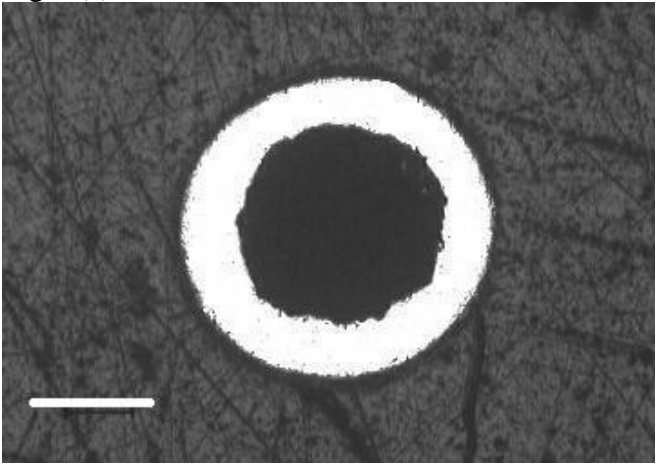

Fig. 2 (b)

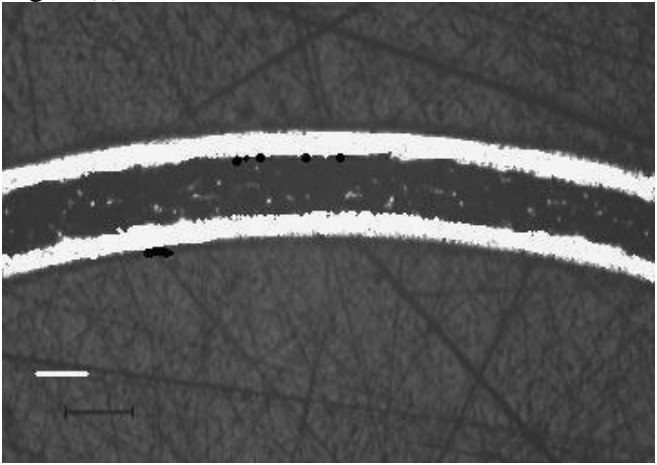

Fig 3

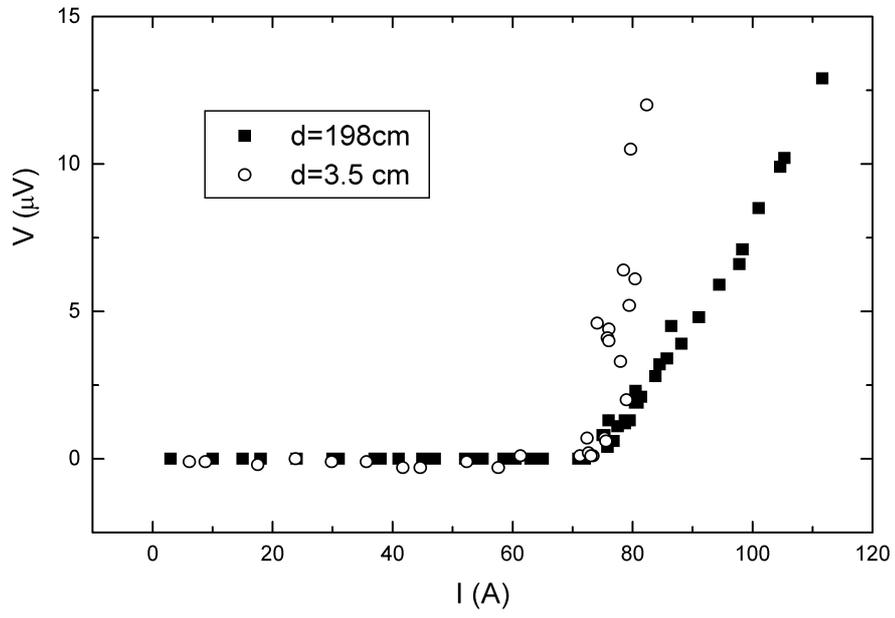